\documentclass{article}
\usepackage{spconf,amsmath,graphicx}
\usepackage{longtable}

\newlength{\secvup}
\title{Improved singing voice separation with chromagram-based pitch-aware remixing}
\name{\begin{tabular}{c}Siyuan Yuan$^{\sharp *}$, Zhepei Wang$^{\flat *}$, Umut Isik$^{\dagger}$, Ritwik Giri$^{\dagger}$ \\
Jean-Marc Valin$^{\dagger}$, Michael M. Goodwin$^{\dagger}$, and Arvindh Krishnaswamy$^{\dagger}$\end{tabular}}

\address{$^{\dagger}$ Amazon Web Services \\
$^{\sharp}$ Stanford University $^{\flat}$ University of Illinois at Urbana-Champaign}

\begin{document}

\maketitle

\begin{abstract}
Singing voice separation aims to separate music into vocals and accompaniment components. One of the major constraints for the task is the limited amount of training data with separated vocals. Data augmentation techniques such as random source mixing have been shown to make better use of existing data and mildly improve model performance. We propose a novel data augmentation technique, chromagram-based pitch-aware remixing, where music segments with high pitch alignment are mixed. By performing controlled experiments in both supervised and semi-supervised settings, we demonstrate that training models with pitch-aware remixing significantly improves the test signal-to-distortion ratio (SDR).
\end{abstract}
\section{Introduction}
\label{sec:intro}

Singing voice separation is the task of separating vocals from music. It is often a crucial first step for many applications including music editing, singer identification, lyrics alignment, and transcription, singing voice synthesis training, and tone analysis. Recent work has primarily focused on using various deep neural network architectures in a supervised manner \cite{Nugraha2016, Uhlich2017, Huang2015, Stoter2019, Takahashi2017, Defossez2019}; training on music libraries with paired vocal and accompaniment as ground truth.

% Spectrogram-based models take spectrograms of audio mixtures as inputs and spectrograms of sources as targets \cite{Nugraha2016, Uhlich2017, Huang2015, Stoter2019, Takahashi2017}. On the other hand, several end-to-end time-domain methods have been proposed. A canonical example is Wave-unet \cite{Stoller2018}, a one-dimensional adaption of the U-Net architecture separating sources directly in the time domain. Other time-domain models include ConvTasNet \cite{Luo2019} and Demucs \cite{Defossez2019}. To aggregate the advantages of spectrogram and time-domain systems, \cite{Song2021} propose CatNet that consists a UNet branch with spectrogram as input and a WavUNet branch using waveform as input. The current best network on MUSDB-18 is \cite{Kong2021} a deep residual U-Net with 143 layers to estimate decoupled magnitude and phase.

Despite significant progress, the bottleneck of further improving the performance of supervised models is primarily the lack of music libraries with isolated sources as ground truth labels. Multi-track datasets that are publicly available for singing voice separation, like MIR-1K \cite{Hsu2021}, ccMixter \cite{Liutkus2014}, and MUSDB \cite{Rafii2017}, are limited to just hours of audio. This limited amount of training data constrains the use of larger networks due to overfitting issues. Artificially increasing the size and variety of the datasets through data augmentation presents an opportunity to enhance large models' ability to generalize. Previous augmentation methods include remixing audio recordings, swapping left and right channels, shifting pitches, and scaling and stretching audio recordings \cite{Pretet2019, Uhlich2017, CohenHadria2019}. However, these methods, individually or combined, have been shown empirically to enhance the model performance only by a marginal amount \cite{Pretet2019}. The work \cite{Song2021}  proposes a data augmentation method, mix-audio augmentation, that randomly mix audio segments from the same source. Although it was shown to be effective, the improvement over random mixing is still limited.

This paper introduces a novel data augmentation method, \emph{chromagram-based pitch-aware remixing}, where sources in song segments with similar pitch content are mixed producing mixtures that are more ``realistic" than random mixing, and more diverse than mix-audio augmentation outputs. \cite{Chiuetal} had experimented with a similar idea for violin/piano, but with a thresholding of match scores to select segments. We instead use the softmax of the match scores as probabilities of selection, and use a temperature parameter, $T$, to adjust the diversity of the mixing from ``realistic" to ``random". To show the effectiveness of our technique, we compare pitch-aware mixing in a supervised setting with the random mixing and mix-audio augmentations. Benchmarking on the MUSDB-18 test set, we show that pitch-aware remixing is significantly more effective. We also find that the best results come from setting the temperature component at a non-zero value; meaning that it is beneficial to remix songs that match well in pitch, but not in a way that limits the diversity of remixes. 

Besides data augmentation, noisy self-training \cite{xie2019} is another direction to compensate for the limited amount of labeled data by utilizing a large amount of publicly available unseparated, or separated but noisy data by self-labeling them using a teacher model. In this setting, the quality of the initial teacher model tends to be important for starting the bootstrapping process. We show the effectiveness of chromagram-based pitch-aware remixing also in a student-teacher setting. We add the chromagram-based remixing to the workflow of \cite{Wang2021}, where we use the above-discussed supervised-trained model as teacher to significantly improve on the student baseline. We obtain further gains by incorporating chromagram-based remixing into the student-training.

% Training a student model with pitch-aware mixing on the pseudo-labels assigned by  teacher, we obtain a chroma student model that reaches the SOTA on MUSDB-18. The contribution of this work is listed as follows:

%\begin{itemize}
%\item We propose a chromagram-based mixing strategy to increase the variety of training samples that sound more "realistic" than purely random mixing

%\item We show that using the new mixing strategy alone empirically leads to evident performance boost for supervised teacher model training; By combining chroma-mixing with a teacher-student semi-supervised framework, we reach the SOTA in terms of test SDR on MUSDB-18.
%\end{itemize}

\vspace{\secvup}
\section{Proposed Method}
In this section, we describe our chromagram-based pitch-aware augmentation strategy, and review the teacher-student framework in \cite{Wang2021}.

\vspace{\secvup}
\subsection{Chromagram-based Pitch-aware Remixing}

\subsubsection{Chromagram}
Chromagram or chroma-based features are a widely used and powerful technique for music alignment and synchronization. Chromagram is closely related to the twelve different pitch classes. The main idea is to aggregate each pitch class across octaves for a given local time window to obtain a 1-D vector expressing how the representation's pitch content within the time window is spread over the twelve chroma bands. Shifting the time window across the music results in a 2-D Time-Chroma representation. We leverage chromagram correlation between song segments as a metric to quantify song similarities because of its high robustness to variations in timbre and closely connected to the musical harmony.

\vspace{\secvup}
\subsubsection{Incorporating pitch-aware re-mixing into network training}
\begin{figure}[htb]

\begin{minipage}[b]{1.0\linewidth}
  \centering
  \centerline{\includegraphics[width=9cm]{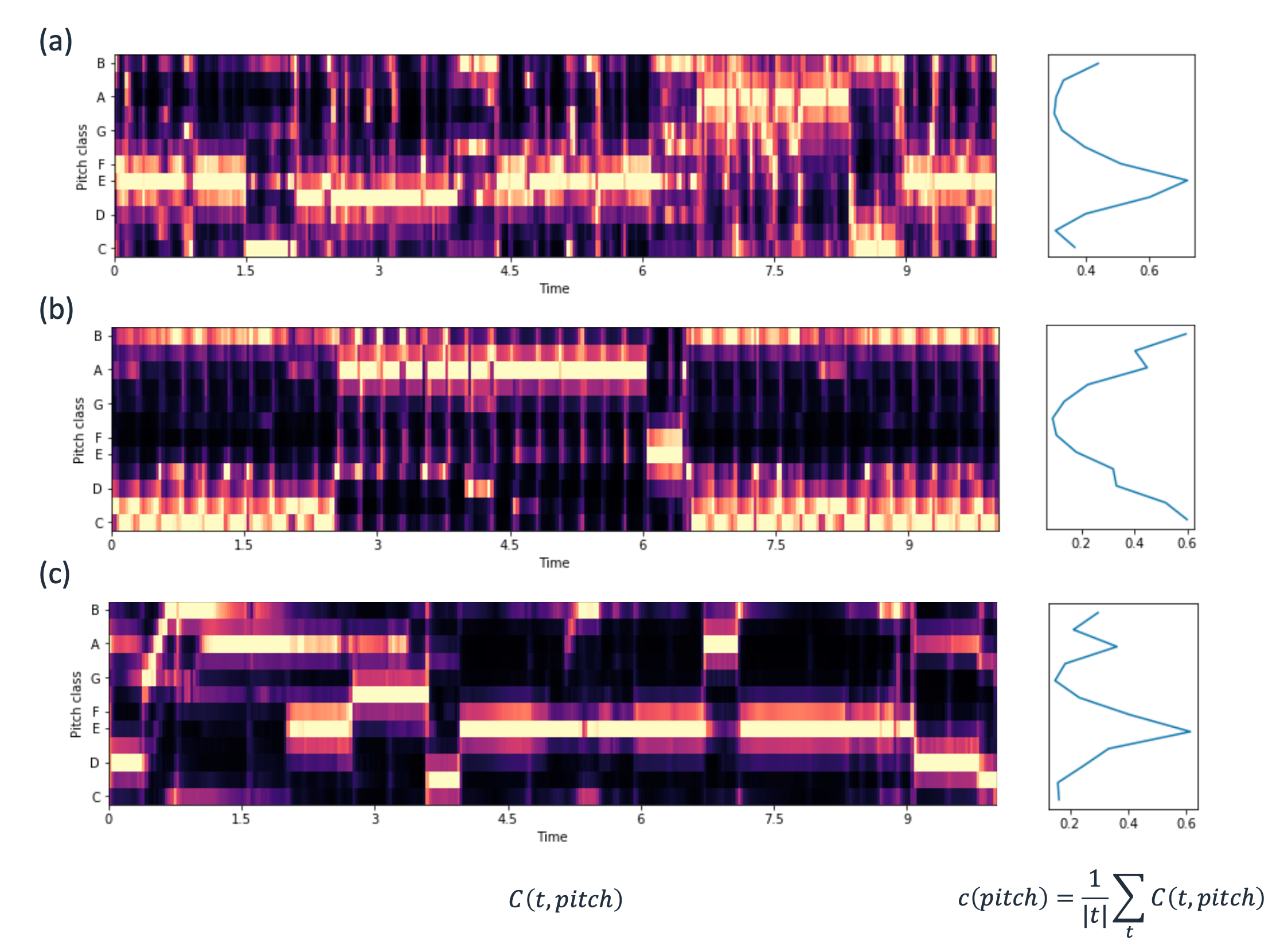}}
%  \vspace{2.0cm}
\end{minipage}
\caption{Accompaniment chromagram $C$ and chroma vectors $c$ of three randomly chosen 10-second segments from MUSDB. Normalized chroma-vector cross-correlation scores, $s_{ab}=0.88$, $s_{ac}=0.98$, indicates segments (a) and (c) share more similarities than (a) and (b).}
\label{fig:chroma}
\end{figure}

We perform chromagram matching of music segments on the fly. For each input mixture, $x_0$, we compute its vocal (or accompaniment) 2-D chromagram, $C(t,\mathrm{pitch})$ using a python package, \texttt{librosa}, and then take the average along time, $t$, to obtain a twelve-dimensional chroma vector, $c_0(\mathrm{pitch})$. We then load $n$ random $t$-second song segments from the same dataset, and compute their vocal or accompaniment chroma vectors, $c_j$, where $j=1...n$. By performing normalized cross-correlation between $c_0$ and $c_j$, we obtain $n$ scores, $s_{0j}$, where $j=1...n$. Song segments having similar pitch content to $x_0$ would have a higher score. Taking softmax with the temperature of $T$ on $(s_{0j})_{j=1}^{n}$, we obtain a probability distribution $(p_j)_{j=1}^{n}$, from which we draw an index $j'$, and conduct source mixing to obtain a new mixture, $\Tilde{x_0}=x_{0,voc}+x_{j',acc}$. New mixtures are more likely to be obtained by mixing song segments with identical pitch content if we lower the softmax temperature. With higher temperature, the song segments to be mixed are more likely to be randomly chosen. Figure \ref{fig:chroma} show examples of chromagram and chroma vectors for three randomly chosen 10-second accompaniment segments from MUSDB. $s_{ac}=0.98 > s_{ab}=0.88$. Therefore, segments, (a) and (c), are more likely to remix due to the similar chromagram features. \\

Note that there are two options to compute chroma vectors of the song segments by using either vocal or accompaniment. Using vocal would mean that we rely on the vocal-to-vocal (voc2voc) match to quantify segment similarity, otherwise we rely on accompaniment-to-accompaniment (acc2acc) match. We experiment with both options.

\vspace{\secvup}
\subsection{Noisy self-training framework}
The framework consists of the following steps:
\begin{enumerate}
\itemsep0em 
    \item Train a teacher separator network $M_0$ on a small labeled dataset $D_l$.
    \item Assign pseudo-labels for the large unlabeled dataset $D_u$ with $M_0$ to obtain the self-labeled dataset $D_0$.
    \item Filter data samples with a pretrained voice activity detector (VAD) from $D_0$ to obtain $D_{f0}$.
    \item Train a student network $M_1$ with $D_l\cup D_{f0}$.
    
\end{enumerate}

\vspace{\secvup}
\subsection{Separator Network}
We adopt the same PoCoNet \cite{Isik2020} architectures for both teacher and student models as \cite{Wang2021}. The inputs are the concatenation of real and imaginary parts of the mixture’s STFT spectrogram. The output is the complex ratio masks for each source. The wave-form signal is obtained by applying inverse STFT transform on the estimated spectrograms. \\
The separator is a fully-convolutional 2D U-Net architecture with DenseNet and attention blocks. Each DenseNet block contains three convolutional layers, each followed by batch normalization and Rectified Linear Unit (ReLU). Convolutional operations are causal in the time direction but not in the frequency direction. We choose a kernel size of 3 × 3 and a stride size of 1, and the number of channels increases from 32, 64, 128 to 256. We control the size of the network by varying the number of levels in U-Net and the maximum number of channels. In the attention module, the number of channels is set to 5 and the encoding dimension for key and query is 20. Frequency-positional embeddings are applied to each time-frequency bin of the input spectrogram.

\vspace{\secvup}
\subsubsection{Loss function}
For each output source, the loss function is the weighted sum of waveform and spectral loss:
\begin{equation} \label{eq1}
\mathcal{L}_s(y, \hat{y})=\lambda_{\mathrm{audio}}\mathcal{L}_{\mathrm{audio}}(y,\hat{y})+\lambda_{\mathrm{spec}}\mathcal{L}_{\mathrm{spec}}(Y, \hat{Y})
\end{equation}
where $s$ is the output source $\mathrm{voc}$ or $\mathrm{acc}$, $y$ and $\hat{y}$ are time domain groundtruth and the network output. $Y$ and $\hat{Y}$ are the corresponding STFT magnitude spectrograms. We choose both $L_{audio}(\cdot)$ and $L_{spectral}(\cdot)$ to be $l1$ loss. The total loss is the weighted sum of the two sources:
\begin{equation} \label{eq2}
\mathcal{L}(y,\hat{y})=\lambda_{\mathrm{voc}}\mathcal{L}_{\mathrm{voc}}(y, \hat{y})+\lambda_{\mathrm{acc}}\mathcal{L}_{\mathrm{acc}}(Y,\hat{Y})
\end{equation}

\vspace{\secvup}
\section{Experimental setup}
\subsection{Training Dataset}
Following \cite{Wang2021}, we use MIR-1K \cite{Hsu2021}, ccMixter \cite{Liutkus2014}, and the training partition of MUSDB \cite{Rafii2017} as the labeled dataset for supervised training with roughly 11 hours of recordings. To train the student model with a large unlabeled dataset with more than 300 hours of recordings from a karaoke app. We train and test at 16kHz. For preprocessing, we compute the STFT spectrograms with a DFT size of 1024 and a hop size of 256.

\subsection{Noisy self-training with pitch-aware mixing}
For both the supervised (teacher) training and student training, we minimize Equation \ref{eq2} with an Adam optimizer with an initial learning rate of $10^{-4}$, and we decrease the learning rate by half for every 100k iterations until it’s no greater than $10^{-6}$. We set $\lambda_{audio} = \lambda_{spec} = 1$ in Equation \ref{eq1} and $\lambda_{voc} = \lambda_{acc} = 1$ in Equation \ref{eq2}. We set the input window size to 10 seconds and the batch size to 1 following the optimal configuration in \cite{Wang2021}.

\vspace{\secvup}
\subsubsection{Supervised (teacher) training}
The prior work \cite{Wang2021} conducts supervised (teacher) training experiments using random mixing for data augmentation, and shows that random mixing with probability of 1 leads to the best teacher model. Here, we experiment with two additional data augmentation strategies: mix-audio augmentation \cite{Kong2021} and the proposed pitch-aware mixing. For a fair comparison, the model architectures and hyper-parameters are the same as the best random mixing models in \cite{Wang2021}. We implemented the chromagram-based pitch-aware mixing based on the description in Section 2.1.2. For each training sample, we load 8 other random 10-second segments in the same dataset to compute matching scores. We experiment with both voc2voc and acc2acc chromagram matching. We also experiment with the softmax temperatures $T \in \{ 0,\,0.33,\,1,\,3 \}$. For comparison, we also experiment with the mix-audio strategy: for each 10-second training samples, we randomly select another 10-second segment from the same song, and mix the sources.

\vspace{\secvup}
\subsubsection{Student training}
Student model is trained on both labeled data and DAMP dataset self-labeled by the chroma teacher model. We integrate pitch-aware mixing on student training, and experiment with $T \in \{ 0,\,0.33,\,1,\,3 \}$.

\label{sec:format}

\vspace{\secvup}
\section{Evaluation results and discussion}

\subsection{Evaluation Framework}
Following the SiSEC separation campaign \cite{Stoter2018}, and for the purpose of continuity with other works (c.f. Table \ref{table:other}), we use Signal-to-Distortion Ratio (SDR) to evaluate the separation performance, computed using the python package \texttt{museval}, which partitions each of the 50 songs from the test partition of MUSDB test set into non-overlapping ten-second segments, and takes the median of segment-wise SDR for each song and reports the median from all 50 songs. Using a standalone validation set to choose $T$ would strengthen our arguments; however, given the limited size of the training set, the knowledge gains from using a validation split would be offset by the reduction in the quality of the training set.             

\vspace{\secvup}
\subsection{Supervised (teacher) Model Performance}
The model architecture and hyperparameters we use here are consistent with the best teacher model in \cite{Wang2021}. Instead of using random mixing, we experiment with the pitch-aware mixing using different softmax temperatures and experiment with voc2voc versus acc2acc matching. Table \ref{table:teacher}  shows the test SDR for the experiments. We can see that both mix-audio augmentation and pitch-aware mixing outperform the random mixing baseline, indicating the effectiveness of both our approach and the mix-audio augmentation. It is also clear that pitch-aware mixing outperforms mix-audio in that all chroma teachers outperform the mix-audio augmentation except the high-temperature chroma teacher ($T=3$). We observe that for the chroma teachers (acc2acc), SDR increases by 0.62 dB as $T$ decreasing from 3 to 0.33. High temperatures make chromagram-mixing closer to random mixing. So, higher $T$ leading to poorer performance is in line with the random mixing results; which can be interpreted as diverse but less ``realistic'' training data limiting model performance. However, we see that decreasing $T$ from 0.33 to 0 doesn't improve the model further, which  can be explained with $T=0$, always mixing the best matching songs, causing a less diverse dataset. With $T=0.33$ seems to reach a balanced state obtaining a both diverse and ``realistic'' dataset that leads to the best teacher model improving the average SDR by 1.05 dB compared to the baseline teacher model with random mixing. Our best supervised model even surpasses the student baseline by 0.48 dB.
\begin{table}
\caption{Test performance metrics (SDR in dB) for teacher models. $T$ refers to the softmax temperature, and `voc'/`acc' corresponds to voc2voc/acc2acc matching strategy in section 2.1.2. The best performance is highlighted in bold.}
\vspace{0.08in}
\begin{tabular}{ |p{4.4cm}||p{0.8cm}|p{0.8cm}|p{0.8cm}|  }
 \hline
 Experiments& SDR (V) &SDR (A)&Mean\\
 \hline
Teacher \cite{Wang2021} (Random mixing; i.e. $T=\infty$)   & 6.91    &13.66&   10.29\\
Student \cite{Wang2021} (Random mixing) &   7.8  & 13.92   &10.86\\
 \hline
  Teacher + mix-audio aug&   7.48 & 14.06 & 10.77 \\
 Chroma Teacher (voc; T=1)&   7.76 & 14.02   &10.89 \\ 
 Chroma Teacher (acc; T=3)&   7.57 & 13.88   &10.72 \\
  Chroma Teacher (acc; T=1)&   7.75 & 14.08   &10.92 \\
 Chroma Teacher (acc; T=0.33)&   \textbf{7.92} & \textbf{14.77} & \textbf{11.34} \\
 Chroma Teacher (acc; T=0)&   7.79 & 14.31 & 11.05 \\
 \hline
\end{tabular}
\label{table:teacher}
\end{table}

\vspace{\secvup}
\subsection{Student Model Performance}
We labeled the mixtures from the unlabeled DAMP dataset using the supervised chroma teacher models (acc2acc) with temperatures of 1 and 0.33, respectively. Correspondingly, we train two student models with pitch-aware mixing using temperatures of 1 and 0.33. Test results are shown in Table \ref{table:student}. We can see that our chroma student outperforms the student baseline with random mixing by 0.69 dB. The lower temperature student model performs the best, implying similar conclusions on teacher training that low temperature training results in both diverse and more realistic dataset leading to higher performance boost.
\begin{table}
\caption{Test performance metrics (SDR in dB) for student models.}
\vspace{0.08in}
\begin{tabular}{ |p{4.4cm}||p{0.8cm}|p{0.8cm}|p{0.8cm}|  }
 \hline
 Experiments& SDR (V) &SDR (A)&Mean\\
 \hline
Best student model in \cite{Wang2021}&   7.8  & 13.92   &10.86\\
 \hline
 Chroma Student (T=1)&   8.55 & 14.67   &11.61 \\
 Chroma Student (T=0.33) &     \textbf{8.39}&  \textbf{15.0}   & \textbf{11.70} \\
 
 \hline
\end{tabular}

\label{table:student}
\end{table}

\vspace{\secvup}
\subsection{Comparisons with Other Models}
In Table \ref{table:other}, we compare our best model with other models. We can see that our chroma student model achieves the second highest mean SDR score, indicating the effectiveness of out augmentation approach. Although, the recent ResUnetDecouple+ performs the best, the data augmentation approach is mix-audio augmentation, which is same as in the previous work it builds upon CatNet\cite{Song2021} with lower SDR than we obtain here. The improvement of ResUnetDecouple+ over CatNet is mainly due to the innovative model architecture and decoupling of the magnitude and phase. Therefore, we believe that applying our data augmentation to the ResUnetDecouple+ architecture has the potential for further improvements.

\begin{table}
\caption{Comparison of the proposed method and other baseline models. \cite{Kong2021} is a follow-up work on \cite{Song2021} with the same mix-audio augmentation but with architecture improvement. Its contribution is orthogonal to our data method. Therefore, considering our improvement over \cite{Wang2021}, it is worth experimenting with combining chromagram-based remixing and the ResUNetDecouple+ architecture.}
\vspace{0.08in}
\begin{tabular}{ |p{4.4cm}||p{0.8cm}|p{0.8cm}|p{0.8cm}|  }
 \hline
Name& SDR (V) &SDR (A)&Mean\\
 \hline
Demucs \cite{Defossez2019}&   7.05  & N/A   &N/A\\
MMDenseLSTM\cite{Takahashi2018}&   4.94 & 16.4   & 10.67 \\ 
MT U-Net\cite{Kadandale2020}&   5.28 & 13.04   &9.16 \\
Wang \emph{et. al.} \cite{Wang2021}& 7.8&13.92&10.86 \\
 CatNet\cite{Song2021}&    7.54& 15.18   &11.36 \\
 ResUNetDecouple+\cite{Kong2021}&    \textbf{8.98}& \textbf{16.63}   &\textbf{12.81} \\
 \hline
  Ours(chroma student)&    8.39& 15.0   &11.70 \\
  \hline
\end{tabular}
\label{table:other}
\end{table}

\vspace{\secvup}
\subsection{Discussion}
The experimental results show that our pitch-aware mixing demonstrates noticeable improvement over the random mixing baseline and mix-audio augmentation. A potential problem with random mixing is that it produces a considerable amount of mixtures by songs with dramatically different pitch content. These mixtures could sound ``unrealistic'' and are more likely to incur domain shift to the underlying distribution of real songs. Besides, these samples could be easier than real songs to separate considering that vocal and accompaniment contain mismatch spectral patterns. These ``unrealistic'' training data could make it difficult for the separator to perform well on unseen realistic test data. While the mix-audio strategy is better than random mixing, the training samples generated are not as diverse as our method, since mix-audio only considers segments from the same song to form mixtures. In contrast, with a larger sample space as well as the temperature component, the proposed method is able to better adjust the trade-off between being diverse and realistic.                   
\vspace{\secvup}
\section{Conclusion}
We introduced a novel data augmentation strategy for singing voice separation. Our approach is chromagram-based and pitch-aware; aiming to mix song segments with similar pitch content to form mixtures that are more likely to resemble real songs while maintaining diversity. Experimental results on the noisy-self training framework show that pitch-aware mixing improves model training compared to random mixing and mix-audio augmentation. Future work would be investigating the effectiveness of our approach on other architectures and other music-related tasks.

% Below is an example of how to insert images. Delete the ``\vspace'' line,
% uncomment the preceding line ``\centerline...'' and replace ``imageX.ps''
% with a suitable PostScript file name.
% -------------------------------------------------------------------------
% \begin{figure}[htb]

% \begin{minipage}[b]{1.0\linewidth}
%   \centering
%   \centerline{\includegraphics[width=8.5cm]{image1}}
% %  \vspace{2.0cm}
%   \centerline{(a) Result 1}\medskip
% \end{minipage}
% %
% \begin{minipage}[b]{.48\linewidth}
%   \centering
%   \centerline{\includegraphics[width=4.0cm]{image3}}
% %  \vspace{1.5cm}
%   \centerline{(b) Results 3}\medskip
% \end{minipage}
% \hfill
% \begin{minipage}[b]{0.48\linewidth}
%   \centering
%   \centerline{\includegraphics[width=4.0cm]{image4}}
% %  \vspace{1.5cm}
%   \centerline{(c) Result 4}\medskip
% \end{minipage}
% %
% \caption{Example of placing a figure with experimental results.}
% \label{fig:res}
% %
% \end{figure}

%\vfill
%\pagebreak

\bibliographystyle{IEEEbib}
\bibliography{strings,refs}

\end{document}